\documentclass{PoS}

\usepackage{dsfont}
\usepackage{amsmath}
\usepackage{subfig}

\DeclareMathOperator{\Tr}{Tr}
\DeclareMathOperator{\R}{Re}

\title{Dualization of non-abelian lattice gauge theory with Abelian Color Cycles (ACC)}

\ShortTitle{Abelian color cycles}

\author{\speaker{Carlotta Marchis}%
	\thanks{This work is supported by the FWF DK W1203 
		{\sl ''Hadrons in Vacuum, Nuclei and Stars''}, and partly also by the FWF Grant.\ Nr.\ I 1452-N27, as well as the
		DFG TR55, {\sl ''Hadron Properties from Lattice QCD''}.}\\
	University of Graz\\
	E-mail: \email{carla.marchis@uni-graz.at}}

\author{Christof Gattringer\\
        University of Graz\\
        E-mail: \email{christof.gattringer@uni-graz.at}}

\abstract{We discuss a new approach to strong coupling expansion and dual representations for non-abelian lattice gauge theories. 
The Wilson gauge action is decomposed into a sum over "abelian color cycles" (ACC), which are loops around plaquettes visiting 
different colors at the corners. ACCs are complex numbers and thus commute such that a dual representation of a non-abelian 
theory can be obtained as in the abelian case. We apply the ACC approach to SU(2) and SU(3) lattice gauge theory and exactly 
rewrite the two partition sums in a strong coupling series where all gauge integrals are known in closed form.}

\FullConference{34th annual International Symposium on Lattice Field Theory\\
	24-30 July 2016\\
	University of Southampton, UK}

\begin{document}
	
\section{Introduction}
Exactly rewriting lattice field theories in terms of new, so-called "dual variables" is a strategy that has been developed and used in 
recent years to overcome complex action problems for lattice field theories at finite density. The Boltzmann factor is decomposed into
local factors which are then expanded, such that subsequently the original degrees of freedom can be integrated out in closed form. 
The partition function turns into a sum over configurations of the expansion indices which constitute the dual variables. 
For several models it was found that this strategy leads to a representation of the partition function with only real and positive 
weights, such that a Monte Carlo simulation in terms of the dual variables solves the complex action problem.    

However, mapping lattice field theories to a dual representation is interesting beyond a possible application for finite
density simulations. The dual variables have to obey constraints which give rise to an interesting geometrical interpretation 
of the dual variables: The gauge field degrees of freedom are described by surfaces that can either be closed surfaces or are 
bounded by loops that represent the matter fields (see, e.g., the reviews 
\cite{Chandrasekharan:2008gp,deForcrand:2010ys,Gattringer:2014nxa}). 
The structure of the constraints and thus the geometrical structure of the dual degrees of freedom 
is of course determined by the symmetries of the theory in the conventional representation. While for U(1) gauge fields 
the geometrical structure is simple and well understood (see, e.g., \cite{Mercado:2013ola}), for non-abelian gauge theories no
clear picture has emerged yet (for different non-abelian dualization strategies see the references in \cite{Gattringer:2016lml}).
 
In this contribution we discuss a new approach for the dualization of non-abelian lattice gauge theories, where the traces
in the Wilson plaquette action are decomposed into color sums over colored loops around plaquettes, so-called Abelian color
cycles (ACC). The ACCs commute such that the same dualization strategy as in the U(1) case can be applied. We present the 
ACC approach for the gauge groups SU(2) and SU(3) and discuss the corresponding constraints.

\section{ACC dualization for SU(2) lattice gauge theory}

\noindent
The Wilson action for SU(2) lattice gauge theory reads:
\begin{equation}
S_G[U] \; = \; -\dfrac{\beta}{2} \sum_{x,\mu < \nu} \!\!
\Tr \; U_{x,\mu} \, U_{x+\hat{\mu},\nu} \, U_{x+\hat{\nu},\mu}^{\dagger} \, U_{x,\nu}^\dagger \; ,
\label{eq:actionsu2}
\end{equation}
where $U_{x,\mu} \in$ SU(2), living on the links of a 4-dimensional lattice, are the dynamical degrees of freedom of the theory.
We decompose the action into a sum over ACCs, by explicitly writing the color sums for trace and the matrix products,
\begin{equation}
\label{eq:abelianactionsu2}
S_G[U] \; = \; -\dfrac{\beta}{2} \sum_{x,\mu < \nu} \; \sum_{a,b,c,d=1}^{2} 
U_{x,\mu}^{ab} \, U_{x+\hat{\mu},\nu}^{bc} \, U_{x+\hat{\nu},\mu}^{dc \ \star} \, U_{x,\nu}^{ad \ \star} \; .
\end{equation} 
The ACCs are the products $U_{x,\mu}^{ab} U_{x+\hat{\mu},\nu}^{bc} U_{x+\hat{\nu},\mu}^{dc \ \star} U_{x,\nu}^{ad \ \star}$
of the matrix elements of the four link elements and are labelled by 4 color indices $a,b,c$ and $d$. Since each color
index has 2 possible values we have $2^4 = 16$ different ACCs, which are complex numbers and therefore commute with 
each other. 

\begin{figure}[t]
	\begin{center}
		\includegraphics[scale=1.10,clip]{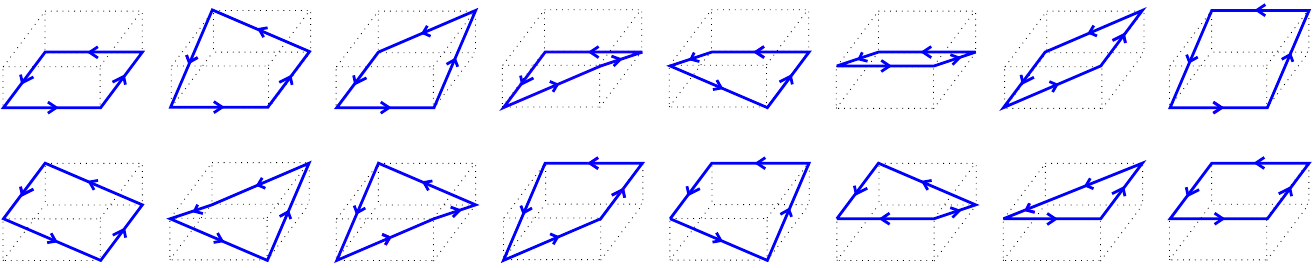}
	\end{center}
	\caption{The 16 possible abelian color cycles attached to a given plaquette. 
	In the dual representation their occupation is given by the corresponding cycle occupation 
	number $p_{x,\mu\nu}^{abcd} \in \mathds{N}_0$.}
	\label{fig:allcycles}	
\end{figure}

It is convenient to introduce a geometrical representation for the link elements as arrows on a 4-dimensional lattice with 2 
layers representing the two possible values of the color indices. More specifically, the element $U_{x,\mu}^{ab}$ is 
represented by an arrow connecting the layer $a$ at site $x$ to the layer $b$ at $x + \hat{\mu}$. Complex conjugation
corresponds to reversing the arrow. With this convention the ACCs 
$U_{x,\mu}^{ab} U_{x+\hat{\mu},\nu}^{bc} U_{x+\hat{\nu},\mu}^{dc \ \star} U_{x,\nu}^{ad \ \star}$ correspond to paths in color space 
closing around plaquettes. In Fig.~\ref{fig:allcycles} we show all 16 cycles that are generated when varying the color labels 
$a,b,c,d$.

Using the ACCs we can now rewrite the partition sum as follows:
\begin{align}
Z  & =  \int \! D[U] \, e^{-S_G[U]} \; = \; \int \! D[U] \prod_{x,\mu<\nu} \prod_{a,b,c,d = 1}^{2} 
e^{\frac{\beta}{2} U_{x,\mu}^{ab} U_{x+\hat{\mu},\nu}^{bc} U_{x+\hat{\nu},\mu}^{dc \ \star} U_{x,\nu}^{ad \ \star}} 
\nonumber \\
& = \int \! D[U] \prod_{x,\mu<\nu} \prod_{a,b,c,d = 1}^{2} \sum_{p_{x,\mu\nu}^{abcd} = 0}^{\infty} 
\dfrac{ \left( \beta/2 \right)^{p_{x,\mu\nu}^{abcd}}}{p_{x,\mu\nu}^{abcd}\, !} 
\left( U_{x,\mu}^{ab} U_{x+\hat{\mu},\nu}^{bc} U_{x+\hat{\nu},\mu}^{dc \ \star} 
U_{x,\nu}^{ad \ \star} \right)^{p_{x,\mu\nu}^{abcd}} 
\nonumber \\
& = \sum_{\{p\}} \left[ \prod_{x,\mu<\nu} \prod_{a,b,c,d}  
\dfrac{ \left(\beta/2 \right)^{p_{x,\mu\nu}^{abcd}}}{p_{x,\mu\nu}^{abcd}\, !} \right] 
\prod_{x,\mu} \int \! \! dU_{x,\mu} \; \prod_{a,b} \left( U_{x,\mu}^{ab} \right) ^{N_{x,\mu}^{ab}} 
\left( U_{x,\mu}^{ab \ \star} \right) ^{\overline{N}_{x,\mu}^{ab}} \; .
\label{eq:partitionsumsu2}
\end{align}
In the first line we rewrite the exponential of the action, which is a sum over plaquettes and over color indices,  
into a product over plaquettes and color indices, such that we obtain individual expontentials for all ACCs.  
In the second step we expand each of these exponentials in a Taylor series, introducing individual expansion coefficients 
$p_{x,\mu\nu}^{abcd} \in \mathds{N}_0$ for each ACC, which we refer to as \textit{cycle occupation numbers}. 
Finally, in the last line we reorganize the factors and introduce the sum over all configurations of 
cycle occupation numbers, $\sum_{\{p\}} = \prod_{x,\mu<\nu} \prod_{a,b,c,d = 1}^{2} \sum_{p_{x,\mu\nu}^{abcd} = 0}^{\infty}$.
After reordering the factors of link elements it is convenient to introduce the exponents for 
$U_{x,\mu}^{ab}$ and  $U_{x,\mu}^{ab\, \star}$ as
$N_{x,\mu}^{ab} \; = \; \sum_{\nu:\mu<\nu}p_{x,\mu\nu}^{abss} + \sum_{\rho:\mu>\rho}p_{x-\hat{\rho},\rho\mu}^{sabs}$ and
$\overline{N}_{x,\mu}^{abcd} \; = \; \sum_{\nu:\mu<\nu}p_{x-\hat{\nu},\mu\nu}^{ssba} + \sum_{\rho:\mu>\rho}p_{x,\rho\mu}^{assb} \; ,$
where the label $s$ stands for the independent summation of the color indices replaced 
by it, e.g., $p_{x,\mu\nu}^{abss} = \sum_{c,d =1}^{2} p_{x,\mu\nu}^{abcd}$.

In order to compute the remaining integrals over the gauge links in the last line of \eqref{eq:partitionsumsu2},
we choose an explicit parametrization of the SU(2) link variables
\begin{equation}
\label{eq:parametrizationsu2}
U_{x,\mu} =\left(
\begin{array}{cc}
\cos\theta_{x,\mu} \, e^{i\alpha_{x,\mu}}  & \sin\theta_{x,\mu} \, e^{i\beta_{x,\mu}}\\
-\sin\theta_{x,\mu} \, e^{-i\beta_{x,\mu}} & \cos\theta_{x,\mu} \, e^{-i\alpha_{x,\mu}}
\end{array} \right) , \ \theta_{x,\mu} \in [0,2\pi], \ \ \alpha_{x,\mu}, \, \beta_{x,\mu} \in [0,\pi/2] \; .
\end{equation}
The corresponding Haar measure reads 
$dU_{x,\mu} = (2 \pi^2)^{-1} \, d\theta_{x,\mu}\sin\theta_{x,\mu} \cos\theta_{x,\mu} \,
d\alpha_{x,\mu} \, d\beta_{x,\mu}$.
All gauge integrals can now be computed in closed form and one finds for the partition sum,
\begin{equation}
Z \; = \; \sum_{\{p\}} W_{\beta}[p] \; W_H[p] \; (-1)^{\sum_{x,\mu}J_{x,\mu}^{21}} \; 
\prod_{x,\mu}  \delta(J_{x,\mu}^{11}-J_{x,\mu}^{22}) \; \delta(J_{x,\mu}^{12}-J_{x,\mu}^{21}) \; ,
\label{eq:partitionsum4}
\end{equation}
where $W_{\beta}[p]$ is the weight factor collecting the coefficients of the Taylor expansion (the term inside the square 
brackets in (\ref{eq:partitionsumsu2})). Evaluating the gauge link integrals in (\ref{eq:partitionsumsu2}) gives
additional weight factors $W_H[p]$ from integrating the $\theta_{x,\mu}$, 
which are related to beta-functions (see \cite{Gattringer:2016lml} for their explicit form). 
Both, $W_{\beta}[p]$ and $W_H[p]$ are real and positive. However, note that the partition 
sum \eqref{eq:partitionsum4} also contains the explicit sign factor $(-1)^{\sum_{x,\mu}J_{x,\mu}^{21}}$ 
which origins from the minus sign in the 2,1 matrix element in the parametrization (\ref{eq:parametrizationsu2}) of our SU(2) link 
variables. The two Kronecker deltas come from the integration over the phases $\alpha_{x,\mu}$ and $\beta_{x,\mu}$ and 
give rise to two constraints on each link. These constraints link together components of the currents $J_{x,\mu}^{ab}$ defined
as   
\begin{equation}
\label{eq:Jfluxes}
J_{x,\mu}^{ab} \; = \; N_{x,\mu}^{ab} - \overline{N}_{x,\mu}^{ab} \; = \; 
\sum_{\nu:\mu<\nu}[\, p_{x,\mu\nu}^{abss} - p_{x-\hat{\nu},\mu\nu}^{ssba} \, ] - 
\sum_{\rho:\mu>\rho}[\, p_{x,\rho\mu}^{assb} - p_{x-\hat{\rho},\rho\mu}^{sabs} \, ] \; .
\end{equation}
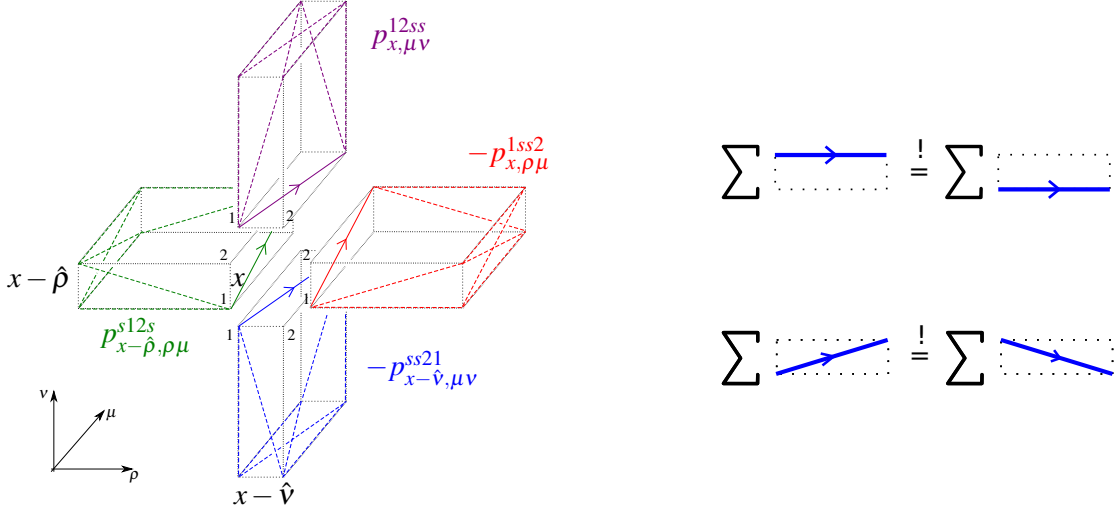
\begin{figure}[t!]
	\centering
	\begin{minipage}[c]{.45\textwidth}
		\centering
		\def\svgwidth{1\textwidth}
\begingroup%
  \makeatletter%
  \providecommand\color[2][]{%
    \errmessage{(Inkscape) Color is used for the text in Inkscape, but the package 'color.sty' is not loaded}%
    \renewcommand\color[2][]{}%
  }%
  \providecommand\transparent[1]{%
    \errmessage{(Inkscape) Transparency is used (non-zero) for the text in Inkscape, but the package 'transparent.sty' is not loaded}%
    \renewcommand\transparent[1]{}%
  }%
  \providecommand\rotatebox[2]{#2}%
  \ifx\svgwidth\undefined%
    \setlength{\unitlength}{613.8236818bp}%
    \ifx\svgscale\undefined%
      \relax%
    \else%
      \setlength{\unitlength}{\unitlength * \real{\svgscale}}%
    \fi%
  \else%
    \setlength{\unitlength}{\svgwidth}%
  \fi%
  \global\let\svgwidth\undefined%
  \global\let\svgscale\undefined%
  \makeatother%
  \begin{picture}(1,0.98331715)%
    \put(0,0){\includegraphics[width=\unitlength]{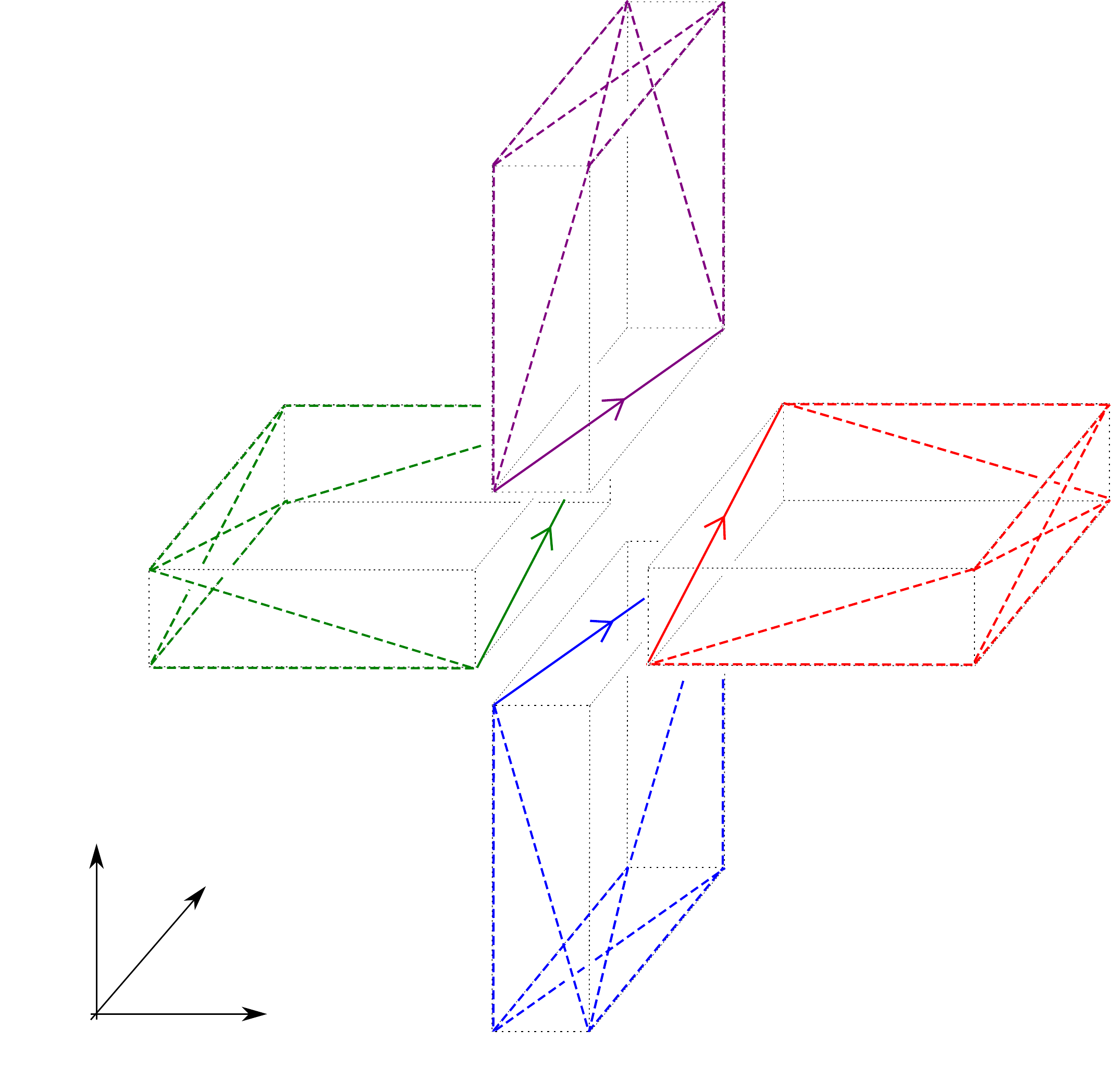}}%
    \put(0.23332398,0.05042547){\color[rgb]{0,0,0}\makebox(0,0)[lb]{\smash{\tiny$\rho$}}}%
    \put(0.05779855,0.20785424){\color[rgb]{0,0,0}\makebox(0,0)[lb]{\smash{\tiny$\nu$}}}%
    \put(0.18809691,0.17134766){\color[rgb]{0,0,0}\makebox(0,0)[lb]{\smash{\tiny$\mu$}}}%
    \put(0.41969012,0.32223267){\color[rgb]{0,0,0}\makebox(0,0)[lb]{\smash{\tiny$1$}}}%
    \put(0.53905742,0.32223267){\color[rgb]{0,0,0}\makebox(0,0)[lb]{\smash{\tiny$2$}}}%
    \put(0.43539796,0.00255061){\color[rgb]{0,0,0}\makebox(0,0)[lb]{\smash{$x - \hat{\nu}$}}}%
    \put(-0.00129821,0.42431637){\color[rgb]{0,0,0}\makebox(0,0)[lb]{\smash{$x - \hat{\rho}$}}}%
    \put(0.88528548,0.66945572){\color[rgb]{1,0,0}\makebox(0,0)[lb]{\smash{$-p^{1ss2}_{x,\rho\mu}$}}}%
    \put(0.68991121,0.25290317){\color[rgb]{0,0,1}\makebox(0,0)[lb]{\smash{$-p^{ss21}_{x-\hat{\nu},\mu\nu}$}}}%
    \put(0.17659314,0.31741349){\color[rgb]{0,0.50196078,0}\makebox(0,0)[lb]{\smash{$p^{s12s}_{x-\hat{\rho},\rho\mu}$}}}%
    \put(0.70004854,0.90077143){\color[rgb]{0.50196078,0,0.50196078}\makebox(0,0)[lb]{\smash{$p^{12ss}_{x,\mu\nu}$}}}%
    \put(0.43094818,0.42708109){\color[rgb]{0,0,0}\makebox(0,0)[lb]{\smash{$x$}}}%
    \put(0.40817043,0.39227248){\color[rgb]{0,0,0}\makebox(0,0)[lb]{\smash{\tiny$1$}}}%
    \put(0.56806396,0.3913509){\color[rgb]{0,0,0}\makebox(0,0)[lb]{\smash{\tiny$1$}}}%
    \put(0.42429803,0.54894049){\color[rgb]{0,0,0}\makebox(0,0)[lb]{\smash{\tiny$1$}}}%
    \put(0.56806396,0.47806051){\color[rgb]{0,0,0}\makebox(0,0)[lb]{\smash{\tiny$2$}}}%
    \put(0.40817043,0.4771389){\color[rgb]{0,0,0}\makebox(0,0)[lb]{\smash{\tiny$2$}}}%
    \put(0.5362927,0.55362977){\color[rgb]{0,0,0}\makebox(0,0)[lb]{\smash{\tiny$2$}}}%
  \end{picture}%
\endgroup%		
	\end{minipage}%
	\hspace{10mm}%
	\begin{minipage}[c]{.45\textwidth}
		\centering
		\hspace*{15mm}\includegraphics[scale=0.8,clip]{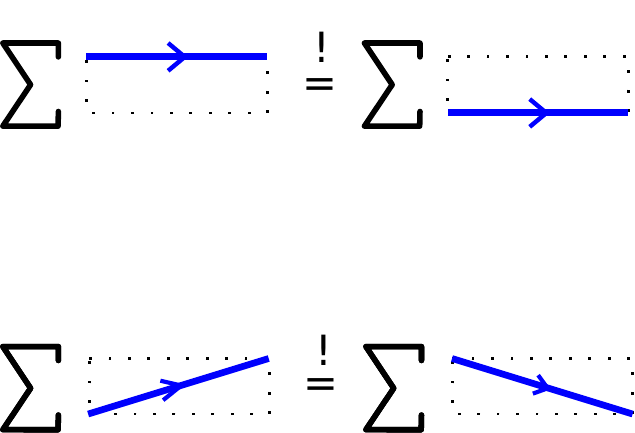}
	\end{minipage}
	\caption{Lhs.: Graphical illustration of the contributions from the cycle occupation numbers to the $J$-flux using the  
		example of the $J_{x,\mu}^{12}$ element. For a description of the plot see the text. 
		Rhs.: Geometrical illustration of the two constraints in Eq.~\protect\eqref{eq:partitionsum4} for the fluxes 
		$J_{x,\mu}^{ab}$ on all links $(x,\mu)$. The first constraint (top) requires the sum over all 1-1 
		fluxes to equal the sum over all 2-2 fluxes. The second constraint (bottom) requires the sum over 1-2 fluxes to equal the 
		sum over 2-1 fluxes.}
	\label{fig:J}
\end{figure}
The $J_{x,\mu}^{ab}$ sum over all cycle occupation numbers that contribute to the flux from color $a$ on site $x$ to color 
$b$ on site $x+\hat{\mu}$. In lhs.\ plot of Fig.~\ref{fig:J} 
we show four of the plaquettes attached to the link $(x,\mu)$ and illustrate how 
they contribute to $J_{x,\mu}^{12}$ as an example. On the link $(x,\mu)$ the flux from color 1 to 2 is kept fixed and represented 
with solid arrows. For every plaquette attached to the link this flux gets contributions from four different cycle occupation numbers, 
which are summed over in the definition \eqref{eq:Jfluxes}, and illustrated with dotted lines.
Thus $J_{x,\mu}^{ab}$ is the total flux from color $a$ on site $x$ to color $b$ on site $x+\hat{\mu}$.

With this interpretation of the $J_{x,\mu}^{ab}$ it is now clear how to interpret the constraints given by the two Kronecker deltas in
\eqref{eq:partitionsum4}: For every link of the lattice, the fluxes on the two color layers have to be equal, and the fluxes between 
the two layers have to match, as represented in the rhs.\ plot of Fig.~\ref{fig:J}. Moreover, the constraints allow for a simple 
interpretation of the sign factor: Since by the constraints the $J_{x,\mu}^{21}$ flux equals the $J_{x,\mu}^{12}$ flux, 
configurations that have an odd number of flux crossings contribute to the partition function with a negative sign. 

In its dual form (\ref{eq:partitionsum4}) the partition function is a sum over configurations of cycle occupation numbers 
$p_{x,\mu\nu}^{abcd} \in \mathbb{N}_{0}$ attached to the plaquettes $(x, \mu < \nu)$. At each link $(x,\mu)$ the 
$p_{x,\mu\nu}^{abcd}$ have to  obey constraints which are expressed in terms of the two Kronecker deltas that relate 
components of the fluxes $J_{x,\mu}^{ab}$ at each link. It is easy to see that a large class of admissible dual pure gauge 
configurations are closed surfaces made of cycle occupation numbers such that at each link the fluxes compensate to 0,
or are such that nontrivial 1-2 fluxes cancel with 2-1 fluxes and 1-1 fluxes with 2-2 fluxes. The latter possibility also allows for 
non-orientable surfaces that are absent in the case of U(1) gauge fields. All these surface configurations have positive signs.

However, configurations with negative sign are not excluded completely by the constraints. We were able 
\cite{Gattringer:2016lml} to construct 
such configurations by stacks of 4 occupied ACCs on a single plaquette, i.e., these configurations appear at ${\cal O}(\beta^4)$.
So far we did not find any other genuine configurations with negative sign that could not be decomposed into factors with
the negative sign 4-stacks among them. This local nature of the negative sign contributions hints at a possible
resummation. 
	
\section{The ACC construction for SU(3)}

\noindent
For SU(3), we follow the same procedure as in SU(2). Starting from the Wilson action:
\begin{equation}
S_G[U] \; = \; -\dfrac{\beta}{3} \sum_{x,\mu < \nu} 
\R  \Tr \, U_{x,\mu} \; U_{x+\hat{\mu},\nu} \, U_{x+\hat{\nu},\mu}^{\dagger} \, U_{x,\nu}^\dagger  \; ,
\label{eq:actionsu3}
\end{equation}
we explicitly write the trace and matrix multiplications as color sums, 
\begin{equation}
\label{eq:abelianactionsu3}
S_G[U] \; = \; -\dfrac{\beta}{6} \sum_{x,\mu < \nu} \; \sum_{a,b,c,d=1}^{3} \left[
U_{x,\mu}^{ab} U_{x+\hat{\mu},\nu}^{bc} U_{x+\hat{\nu},\mu}^{dc \ \star} U_{x,\nu}^{ad \ \star} + 
U_{x,\mu}^{ab \ \star} U_{x+\hat{\mu},\nu}^{bc \ \star} U_{x+\hat{\nu},\mu}^{dc} U_{x,\nu}^{ad} \right] \; .
\end{equation} 
As in the SU(2) case we refer to the products of link matrix elements as ACCs. Note that for SU(3) we explicitly have an ACC and its 
complex conjugate, while in SU(2) there is no such pairing due to the pseudo-reality of SU(2). Again we write the Boltzmann 
factor $e^{-S_G[U]}$ as a product over plaquette coordinates $(x,\mu < \nu)$ and color indices $(a,b,c,d)$ and for each combination 
of indices obtain two Boltzmann factors for the ACC and its complex conjugate. Both are expanded, giving rise to two sets 
of expansion indices $n_{x,\mu\nu}^{abcd} \in \mathds{N}_0$ and  $\bar{n}_{x,\mu\nu}^{abcd} \in \mathds{N}_0$. 
After reorganizing the products
over link matrix elements we find the representation that corresponds to (\ref{eq:partitionsumsu2}) in the SU(2) case,
\begin{equation}
Z = \; \sum_{\{n, \bar{n}\}} \left[ \prod_{x,\mu<\nu} \prod_{a,b,c,d}  
\dfrac{ \left( \beta/6 \right)^{n_{x,\mu\nu}^{abcd} + \bar{n}_{x,\mu\nu}^{abcd}}}{n_{x,\mu\nu}^{abcd} ! \; \; 
\bar{n}_{x,\mu\nu}^{abcd} !} \right] 
\prod_{x,\mu} \int \! \! dU_{x,\mu} \; \prod_{a,b} \left( U_{x,\mu}^{ab} \right) ^{N_{x,\mu}^{ab}} 
\left( U_{x,\mu}^{ab \ \star} \right) ^{\overline{N}_{x,\mu}^{ab}} \; ,
\label{eq:partitionsumsu3}
\end{equation}
where
\begin{gather} 
N_{x,\mu}^{abcd} \; = \; 
\sum_{\nu:\mu<\nu}n_{x,\mu\nu}^{abss} + \bar{n}_{x-\hat{\nu},\mu\nu}^{ssba} + 
\sum_{\rho:\mu>\rho}\bar{n}_{x,\rho\mu}^{assb} + n_{x-\hat{\rho},\rho\mu}^{sabs} \; \; ,
\\
\overline{N}_{x,\mu}^{abcd} \; = \; 
\sum_{\nu:\mu<\nu}\bar{n}_{x,\mu\nu}^{abss} + n_{x-\hat{\nu},\mu\nu}^{ssba} + 
\sum_{\rho:\mu>\rho}n_{x,\rho\mu}^{assb} + \bar{n}_{x-\hat{\rho},\rho\mu}^{sabs} \; \; .
\end{gather} 

For integrating out the SU(3)  gauge links $U_{x,\mu}$ we choose the parametrization \cite{brozan}:
\begin{equation}
\label{eq:parametrizationsu3}
U_{x,\mu} =\left(
\begin{array}{ccc}
c_1 c_2 \, e^{i\phi_1}  \; & s_1 \, e^{i\phi_3} \; & c_1 s_2 \, e^{i\phi_4}\\
s_2 s_3 \, e^{-i\phi_4 -i\phi_5} - s_1 c_2 c_3 \, e^{i\phi_1 +i\phi_2 -i\phi_3} \; & c_1 c_3 \, e^{i\phi_2} \; & - c_2 s_3 \, e^{-i\phi_1 -i\phi_5} - s_1 s_2 c_3 \, e^{i\phi_2 -i\phi_3 +i\phi_4}\\
- s_2 c_3 \, e^{-i\phi_2 -i\phi_4} - s_1 c_2 s_3 \, e^{i\phi_1 -i\phi_2 +i\phi_5} \; & c_1 s_3 \, e^{i\phi_5} \; & c_2 c_3 \, e^{-i\phi_1 -i\phi_2} - s_1 s_2 s_3 \, e^{-i\phi_3 +i\phi_4 +i\phi_5}
\end{array} \right) \; ,
\end{equation}
where $c_i = \cos \theta^{(i)}_{x,\mu}$, $s_i = \sin \theta^{(i)}_{x,\mu}$, with $\theta^{(i)}_{x,\mu} \in [0,\pi/2]$, and $\phi_i = \phi^{(i)}_{x,\mu}$, with $\phi^{(i)}_{x,\mu} \in [0,2\pi]$, and Haar measure 
$dU_{x,\mu}  = (2\pi^5)^{-1} \; d\theta_1 c_1^3 s_1 \; d\theta_2 c_2 s_2 \; d\theta_3 c_3 s_3 \; 
d\phi_1 \; d\phi_2 \; d\phi_3 \; d\phi_4 \; d\phi_5$.
In the following, it will prove convenient to perform the change of variables:
\begin{align}
&n_{x,\mu\nu}^{abcd} - \bar{n}_{x,\mu\nu}^{abcd} = p_{x,\mu\nu}^{abcd} \quad , \quad p_{x,\mu\nu}^{abcd} \in \mathbb{Z} \, , \\
&n_{x,\mu\nu}^{abcd} + \bar{n}_{x,\mu\nu}^{abcd} = |p_{x,\mu\nu}^{abcd}| + 2 \, l_{x,\mu\nu}^{abcd} \quad , \quad l_{x,\mu\nu}^{abcd} \in \mathbb{N}_0 \, ,
\end{align}
and to introduce the fluxes $J_{x,\mu}^{ab} \; = \; N_{x,\mu}^{ab} - \overline{N}_{x,\mu}^{ab}$ 
and $S_{x,\mu}^{ab} \; = \; N_{x,\mu}^{ab} + \overline{N}_{x,\mu}^{ab}$ given explicitly by
\begin{gather}
\label{eq:Jfluxessu3}
J_{x,\mu}^{ab} \; = \! 
\sum_{\nu:\mu<\nu}  [\, p_{x,\mu\nu}^{abss} - p_{x-\hat{\nu},\mu\nu}^{ssba} \, ] - \!\!
\sum_{\rho:\mu>\rho}[\, p_{x,\rho\mu}^{assb} - p_{x-\hat{\rho},\rho\mu}^{sabs} \, ] \; ,\\ \nonumber
S_{x,\mu}^{ab} =  \!\!\!\!
\sum_{\nu:\mu<\nu} \!\!\! [ |p_{x,\mu\nu}^{abss}| + |p_{x-\hat{\nu},\mu\nu}^{ssba}| + 
2(l_{x,\mu\nu}^{abss} + l_{x-\hat{\nu},\mu\nu}^{ssba}) ] 
+ \!\!\!\!
\sum_{\rho:\mu>\rho} \!\!\! [ |p_{x,\rho\mu}^{assb}| - |p_{x-\hat{\rho},\rho\mu}^{sabs}| + 
2(l_{x,\rho\mu}^{assb} - l_{x-\hat{\rho},\rho\mu}^{sabs}) ] \, .
\end{gather}
The geometrical interpretation of the $J_{x,\mu}^{ab}$ is the same as for SU(2), i.e., they represent the total flux from color $a$ on 
site $x$ to color $b$ on site $x+\hat{\mu}$, where now the color indices can be $1,2$ or $3$.

To obtain the final result for the partition sum we substitute the parametrization \eqref{eq:parametrizationsu3} and the Haar measure  
in \eqref{eq:partitionsumsu3}. An additional step is still required in order to be able to perform the Haar measure
integration, because some of the elements $U_{x,\mu}^{ab}$ of the matrix \eqref{eq:parametrizationsu3} are not in the simple form 
$U_{x,\mu}^{ab} = r_{x,\mu}^{ab} e^{i \varphi_{x,\mu}^{ab}}$, 
but are sums $U_{x,\mu}^{ab} = \rho_{x,\mu}^{ab} e^{i \alpha_{x,\mu}^{ab}} + \omega_{x,\mu}^{ab} e^{i \beta_{x,\mu}^{ab}}$. 
For the latter we make use of the binomial theorem $(x + y)^{n} = \sum_{k = 0}^{n} \binom{n}{k}x^{k} y^{n-k}$ 
and rewrite the integrand in \eqref{eq:partitionsumsu3} as
\begin{equation}
\left( U_{x,\mu}^{ab} \right)^{\!N_{x,\mu}^{ab}}  \!
\left( U_{x,\mu}^{ab \ \star} \right) ^{\!\overline{N}_{x,\mu}^{ab}} \; = \!\!
\sum_{m_{x,\mu}^{ab} = 0}^{N_{x,\mu}^{ab}} \, 
\sum_{\overline{m}_{x,\mu}^{ab} = 0}^{\overline{N}_{x,\mu}^{ab}} \!\!
\! \binom{N_{x,\mu}^{ab}}{m_{x,\mu}^{ab}} \! \binom{\overline{N}_{x,\mu}^{ab}}{\overline{m}_{x,\mu}^{ab}} \!
\left( \! \rho_{x,\mu}^{ab} \! \right)^{\!s_{x,\mu}^{ab}} \! \left( \! \omega_{x,\mu}^{ab} \! \right)^{\!S_{x,\mu}^{ab} - s_{x,\mu}^{ab}} 
e^{i\alpha_{x,\mu}^{ab} j_{x,\mu}^{ab}} \, e^{i \beta_{x,\mu}^{ab} \left(J_{x,\mu}^{ab} - j_{x,\mu}^{ab}\right)}.
\end{equation}
This procedure introduces new sets of dual variables, $m_{x,\mu}^{ab}$ and $\overline{m}_{x,\mu}^{ab}$. 
For the sums and differences of these we use the shorthand notation 
$j_{x,\mu}^{ab} = m_{x,\mu}^{ab} - \overline{m}_{x,\mu}^{ab}$, $s_{x,\mu}^{ab} = m_{x,\mu}^{ab} + \overline{m}_{x,\mu}^{ab}$.

\begin{figure}[t]
\begin{center}
		\hspace*{-1mm}
		\includegraphics[scale=0.69,clip]{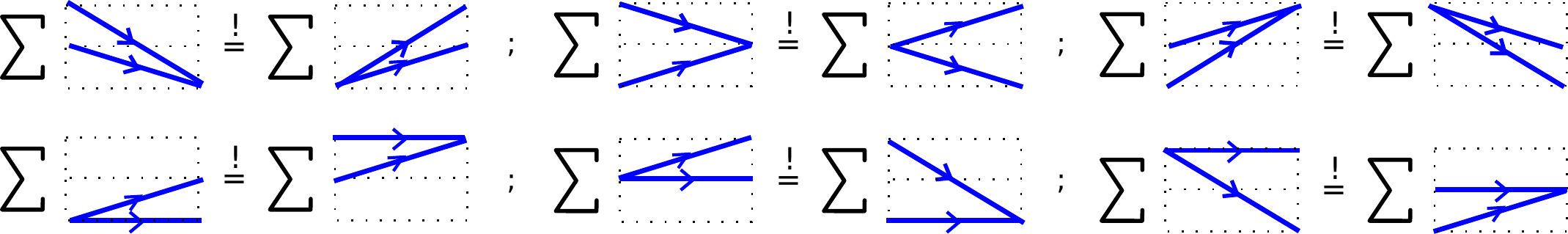}
	\end{center}
	\caption{Geometrical illustration of the constraints in Eq.~(3.11) for the fluxes $J_{x,\mu}^{ab}$	
	on all links $(x,\mu)$.  The constraints in the top row imply that the flux out of a color has to equal the flux into that color. The 
	bottom row of constraints governs the exchange of flux between two colors. In the absence of exchange all three colors must 
	have the same flux. Note that the diagrams are overcomplete and only four of them are independent, corresponding to the 
	four constraints in (3.11). \hfill}
	\label{fig:fluxconservationsu3}
\end{figure}

Inserting the matrix elements from (\ref{eq:parametrizationsu3}) and performing the gauge field integration one finds
\begin{align}
&Z \; = \; \sum_{\{p,l\}} \sum_{\{m,\overline{m}\}} W_{\beta}[p,l] \; W_H[p,l,m,\overline{m}] \; 
(-1)^{\sum_{x,\mu}J_{x,\mu}^{12} + J_{x,\mu}^{23} + J_{x,\mu}^{31} - j_{x,\mu}^{23} - j_{x,\mu}^{31}} \; \prod_{x,\mu}  
\delta(J_{x,\mu}^{11} + J_{x,\mu}^{12} - J_{x,\mu}^{33} - J_{x,\mu}^{23})
\nonumber \\
& \hspace{5mm} \times  \; 
\delta(J_{x,\mu}^{22} + J_{x,\mu}^{12} - J_{x,\mu}^{33} - J_{x,\mu}^{31}) \; 
\delta(J_{x,\mu}^{13} + J_{x,\mu}^{12} - J_{x,\mu}^{31} - J_{x,\mu}^{21}) \; 
\delta(J_{x,\mu}^{32} + J_{x,\mu}^{12} - J_{x,\mu}^{23} - J_{x,\mu}^{21}) \; .
\label{eq:partitionsumsu32}
\end{align}
The partition function is a sum over configurations of the cycle occupation numbers $p_{x,\mu\nu}^{abcd} \in \mathbb{Z}$ and the 
dual variables  $l_{x,\mu\nu}^{abcd} \in \mathbb{N}_{0}$, $m_{x,\mu\nu}^{abcd}$ and $\overline{m}_{x,\mu\nu}^{abcd}$. Each 
configuration comes with the real and positive weight factors $W_{\beta}[p,l]$ and $W_{H}[p,l,m,\overline{m}]$ which collect the 
coefficients of the Taylor expansion and the combinatorial factors from the Haar measure integral. Again we find a sign factor 
$(-1)^{\sum_{x,\mu}J_{x,\mu}^{12} + J_{x,\mu}^{23} + J_{x,\mu}^{31} - j_{x,\mu}^{23} - j_{x,\mu}^{31}}$, which comes from the 
explicit minus signs in (\ref{eq:parametrizationsu3}).
The $p_{x,\mu\nu}^{abcd}$ have to  obey constraints which are expressed in terms of the four Kronecker deltas in
\eqref{eq:partitionsumsu32} that relate components of the fluxes $J_{x,\mu}^{ab}$ at each link. The geometrical interpretation of the
constraints is illustrated in Fig. \ref{fig:fluxconservationsu3} using a straightforward generalization 
of the SU(2) graphical representation.
\section{Concluding remarks}
In this paper we have presented a new method for finding a dual representation for non-abelian lattice gauge theories, based on 
strong coupling expansion. The key ingredient for the success of the dualization is a decomposition of the gauge action in terms of 
abelian color cycles (ACC) which are loops in color space around plaquettes. The ACCs are abelian in nature, i.e., they commute, 
and the dualization proceeds as in the abelian case. The link integration can be performed explicitly 
and all expansion coefficients are known in closed form -- they are simple combinatorial factors.

For the case of SU(2), in \cite{Gattringer:2016lml} we presented an extension of the dualization with ACCs by including staggered 
fermions. A remarkable fact is that in the leading terms of the coupled hopping/strong coupling expansion the minus signs cancel 
such that in this limit also a dual simulation is possible without the aforementioned resummation. The exploratory 
results presented here for SU(3) aim at a first assessment of the structure of constraints to be expected for that group.

An interesting open question is whether the dualization strategy based on ACCs allows for a full dualization in the sense that
new gauge fields are introduced on the dual lattice such that the constraints are automatically fulfilled. While for U(1) lattice field 
theory such a dualization is well known (see, e.g., the review \cite{Savit}), the non-abelian case is less understood. Maybe the 
ACC strategy, which is patterned after the abelian approach, leads to progress towards such a full dualization.

\end{document}